\documentclass[12pt,preprint]{aastex}

\newcommand{\Gm}{G010.62$-$00.38}
\newcommand{\Gq}{G012.02$-$00.03}

\newcommand{\kms}{km\,s$^{-1}$}
\newcommand{\HI}{\hbox{H~{\sc i}}}
\newcommand{\HII}{\hbox{H~{\sc ii}}}
\newcommand{\ie}{i.e.,}

\begin{document}

\title{TRIGONOMETRIC PARALLAXES TO STAR-FORMING REGIONS WITHIN 4 kpc OF THE GALACTIC CENTER}

\author{A. Sanna\altaffilmark{1}, M. J. Reid\altaffilmark{2}, K. M. Menten\altaffilmark{1},
T. M. Dame\altaffilmark{2}, B. Zhang\altaffilmark{1}, M. Sato\altaffilmark{1},
A. Brunthaler\altaffilmark{1}, L. Moscadelli\altaffilmark{3}, K. Immer\altaffilmark{1}}

\email{asanna@mpifr-bonn.mpg.de}
\altaffiltext{1}{Max-Planck-Institut f\"{u}r Radioastronomie, Auf dem H\"{u}gel 69, 53121 Bonn, Germany}
\altaffiltext{2}{Harvard-Smithsonian Center for Astrophysics, 60 Garden Street, Cambridge, MA 02138, USA}
\altaffiltext{3}{INAF, Osservatorio Astrofisico di Arcetri, Largo E. Fermi 5, 50125 Firenze, Italy}

\begin{abstract}

We report four trigonometric parallaxes for high-mass star-forming regions within 4~kpc of the Galactic
center.  These measurements were made with the VLBA as part of the BeSSeL Survey.
By associating these sources kinematically with large-scale features in CO and \HI\ longitude-velocity
diagrams, we begin to outline some major features of the inner Milky Way: the Connecting arm, the near and far
3~kpc arms, and the Norma arm.  The Connecting arm in the first Galactic quadrant lies closer to the
Galactic center than the far 3~kpc arm and is offset by the long-bar's major axis near its leading edge,
supporting the presence of an inner Lindblad resonance.  Assuming the 3~kpc arms are a continuous physical
structure, the relative Galactocentric distance of its near and far sides suggests highly elliptical
streamlines of gas around the bar(s) and a bar corotation radius, $r_{\rm CR} \gtrsim 3.6$~kpc.
At a Galactic longitude near $10\degr$ and a heliocentric distance of about 5~kpc, the near
3~kpc arm and the Norma arm intersect on a face-on view of our Galaxy, while passing at different Galactic
latitudes.  We provide an accurate distance measurement to the W\,31 star-forming complex
of $4.95^{+0.51}_{-0.43}$~kpc from the Sun, which associates it with a bright CO feature belonging to the
near 3~kpc arm.

\end{abstract}

\keywords{astrometry --- Galaxy: fundamental parameters --- Galaxy: kinematics and dynamics
--- masers --- techniques: high angular resolution --- stars: individual: W\,31, G010.62$-$00.38, G010.47$+$00.02}

\section{Introduction}

The so-called Galactic Molecular Ring (GMR) is a ridge of intense emission which dominates the
appearance of extended CO gas in the longitude-velocity ($\ell$-$v$) diagram of our Galaxy
(e.g., \citealt{Dame2001}). This prominent feature marks a region of enhanced molecular density
roughly halfway between the Sun and the Galactic center, at Galactocentric radii (R)
between 4 and 6~kpc, which has been shown to likely trace gas emission from the inner spiral arms
(e.g., \citealt{Dobbs2012}). Toward the Galactic center, the inner edge of the GMR would coincide
with the Scutum-Centaurus arm, which we have recently located at an average Galactocentric radius
of about 5~kpc  (Sato et al., submitted.).
Hereafter, we will refer to the region from the inner edge of the GMR to the Galactic center as the
inner Milky Way.

As part of the Bar and Spiral Structure Legacy (BeSSeL) Survey
\footnote{See the BeSSeL website at the following URL: http://bessel.vlbi-astrometry.org/},
we started a detailed study of the gas distribution and velocity field in the inner Milky Way, via
trigonometric parallaxes and proper motions of masers in high-mass star-forming regions (HMSFRs).
In this paper, we constrain some structures located in between the GMR and the Central Molecular
Zone (CMZ; e.g., \citealt{Morris1996}) in the first Galactic quadrant.  \
These structures include the Connecting arm (e.g., \citealt{Fux1999,Marshall2008}),
the near and far 3~kpc arms (e.g., \citealt{Dame2008}), and the Norma arm as it is traced toward
the Galactic center (e.g., \citealt{Bronfman2000}).
The Connecting arm has received little discussion in the literature
(e.g., \citealt{Fux1999} for a short summary; see also \citealt[and references therein]{Rodriguez2008}).
It has been suggested to possibly appear similar to
dust lanes observed in optical images of external galaxies along the extent of their bars,
roughly \emph{connecting} the nuclear ring with the inner tips of the spiral arms
(e.g., see the Hubble Space Telescope composite view of the prototypical
barred spiral galaxy NGC\,1300, or that of NGC\,1097). These dust lanes, often observed
offset toward the leading edges of the bar structure in barred spirals, have been identified as tracers
of shocks (e.g., \citealt{Roberts1979}; \citealt{Athanassoula1992}).
In the following discussion, we will assume a general picture of the inner Galaxy as depicted in
\citet{Churchwell2009}, which accounts for the presence of two bar-like components: the boxy-bulge
(i.e., the Galactic bar) and the long bar.

In this paper, we present trigonometric parallax measurements of 22~GHz H$_2$O and 12~GHz CH$_3$OH masers
obtained with the Very Long Baseline Array (VLBA) for four sites of massive star formation located
in the inner Galaxy.

\clearpage

\section{OBSERVATIONS}

We conducted multi-epoch VLBA (program BR145)
\footnote{The VLBA is operated by the National Radio Astronomy Observatory (NRAO).
The NRAO is a facility of the National Science Foundation operated under cooperative agreement by
Associated Universities, Inc.} observations of the $6_{16}-5_{23}$ H$_2$O
(rest frequency 22.235079 GHz) and the $2_{0}-3_{-1}$E CH$_3$OH (rest frequency 12.178597 GHz)
maser emission toward the four HMSFRs listed in  Table~\ref{tabobs}.
In order to measure trigonometric parallaxes and proper motions,
we switched rapidly between a maser target and three or four extragalactic continuum sources.
These calibrators were selected from our survey \citep{Immer2011} and
included ICRF sources \citep{Fey2004} with accurate positions ($\pm2$~mas).
We placed four geodetic-like blocks throughout each 7-hour track, in order to measure and
remove atmospheric delays for each antenna. Details about the observational strategy can be found in
\citet[see also \citealt{Sato2010}]{Reid2009a}.
Observation and source information are summarized in Tables~\ref{tabobs} and~\ref{tabsou}.

Four adjacent intermediate frequency (IF) bands, each 8 MHz wide, were recorded in
dual circular polarization; each band was correlated to produce 256 spectral channels.
The channel width of 31.25 kHz corresponds to a spectral resolution of 0.42 and 0.77 \kms\
for the H$_2$O and CH$_3$OH maser transitions, respectively. The third IF band was
centered about the LSR velocity (V$_{\rm LSR}$) of the strongest maser feature
detected in our preparatory survey, as reported in Table~\ref{tabobs} (under program BR145A).
The data were processed with the VLBA DiFX software correlator in Socorro \citep{Deller2007}
using an averaging time of about 1~s, which limited the instantaneous field of view of the
interferometer to about $3''$ and $5''$ for the H$_2$O and CH$_3$OH maser observations, respectively.
Data were reduced with the NRAO Astronomical Image Processing System (AIPS) following the
procedure described in \citet{Reid2009a}, using a ParselTongue
scripting interface \citep{Kettenis2006}. Total-power spectra of the 22.2 and 12.2~GHz maser emission toward
\objectname{G010.47$+$00.02}, \objectname{G010.62--00.38}, \objectname{G012.02--00.03}, and
\objectname{G023.70--00.19} from the first epoch data are shown in Figure~\ref{spectrum}.

\section{RESULTS}\label{distance_results}

We modeled the position offsets of compact maser spots with respect to background sources as a function
of time, in order to determine their parallaxes and proper motions (e.g., \citealt{Reid2009a}
for details). During the fitting procedure, we quantified ``a posteriori'' the systematic errors,
via error-floors added in quadrature to the formal fitting uncertainties,
for the E--W and N--S offsets. These error floors were iteratively adjusted to yield values
of chi-squared per degree of freedom near unity.  Results of the parallax and proper motion fitting
for each maser spot and QSO used are listed in Table~\ref{tabres} and displayed in Figures~\ref{paralg10}
to~\ref{paralg23}.  For each source, individual parallax measurements with multiple maser spots and
QSOs were combined in a final measurement and plotted in the right panel of Figure~\ref{paralg10} to~\ref{paralg23}.
The formal uncertainties of the combined fits were multiplied by $\sqrt{N}$, where N is the number
of maser spots, to account for the possibility of correlated position shifts among
the maser spots. Details on individual parallax fittings are presented in Appendix~\ref{appendix}. In Figure~\ref{positions},
the Galactic locations of our sources are superposed on a schematic
face-on view of the Galaxy (re-scaled for a Sun-Galactic center distance of 8.38~kpc), with a number
of spatial features updated from the recent literature, such as the position of the Galactic (e.g., \citealt{Gerhard2002})
and long \citep{Benjamin2005} bars as well as the streamline model of the 3~kpc arms by \citet{Green2011}.

For each source, the line-of-sight velocity component (V$_{\rm LSR}$) and the eastward and northward motions on
the plane of the sky ($\mu_{\rm x}, \mu_{\rm y}$) give the 3-dimensional velocity vector of the star-forming region
as measured with respect to the equatorial heliocentric reference frame (after adding the standard Solar motion to V$_{\rm LSR}$).
In previous papers, to investigate deviations from a circular rotation about the Galactic center
(i.e., the peculiar motion of the HMSFR), we have moved to a reference frame rotating with the
Galaxy (e.g., \citealt{Reid2009b}).  However, for sources with Galactocentric distances $\rm R < 4$~kpc,
which may move along highly elliptical streamlines under the influence of the bulge/bar potential,
we transform to a reference frame at rest at the Galactic center:
$\rm (U_s^{\rm G.C.}, V_s^{\rm G.C.}, W_s^{\rm G.C.})$.
U$_s^{\rm G.C.}$, V$_s^{\rm G.C.}$, and W$_s^{\rm G.C.}$ are directed toward the Galactic center,
in the direction of Galactic rotation and toward the North Galactic Pole, respectively, at
the location of each source\footnote{With respect to velocity components in a rotating reference frame
$\rm (U_s, V_s, W_s)$, the following relation holds: $\rm U_s^{\rm G.C.}=U_s, V_s^{\rm G.C.}=V_s+\Theta(R), W_s^{\rm G.C.}=W_s$,
where $\rm \Theta(R)$ gives the circular Galactic rotation at the source location (cf. \citealt{Reid2009b}).}.
In Table~\ref{tabinner}, we summarize our results together with those obtained for
the other inner Galaxy sources in the literature. In this calculation, we adopted a current ``best-estimate''
of the Galactic parameters, R$_0 =8.38$~kpc and $\Theta_0=243$~\kms, from the trigonometric parallaxes
measured with maser lines \citep{Reid2013} and the revised \emph{Hipparcos} measurements of the Solar motion
from \citet{Schoenrich2010}.

\subsection{Individual Sources}\label{association}

It is possible to associate each of the present HMSFRs to prominent large-scale features in
CO and \HI\ longitude-velocity diagrams of the inner Galaxy ($|\ell| < 30\degr$).
In Figure~\ref{lvdiagram}, we present a longitude-velocity diagram of extended gas emission from
the Galactic CO survey by \citet{Dame2001} with a number of structures relevant for discussing the
inner few kpc of the Galactic center. The near and far 3~kpc arms appear as outlined by \citet{Dame2008}, as parallel
lanes within the range of longitudes where they can be followed clearly in CO ($-12\degr < \ell < +13\degr$).
Note that the far 3~kpc arm shows only weakly in Figure~\ref{lvdiagram} (see \citealt{Dame2008} for a detailed analysis).
In the first quadrant, the line-of-sight velocity pattern of the near side of the Connecting arm,
indicated in Figure~\ref{lvdiagram} following the analysis by \citet[e.g., his Figure~1 and Section~6.1]{Fux1999},
passes through the peak of the terminal velocity curve at positive longitudes.
While crossing the $\ell$-$v$ pattern of the far 3~kpc arm at a Galactic longitude near $10\degr$, this feature
is better isolated at negative latitudes in both \HI\ and CO (e.g., Figure~4 of \citealt{Marshall2008}).
An approximate locus for the Norma/4~kpc arm in the $\ell$-$v$ diagram
is obtained by assuming a logarithmic spiral with the following constraints:
1) an expanding motion fixed at $\ell = 0\degr$ of $-29.3$~\kms, as obtained from CO
spectra in absorption toward the Galactic center (first noticed by \citealt{Kerr1967} in \HI;
see also \citealt{Greaves1994} for CS absorption);
2) a southern tangent at a Galactic longitude of $-32.5\degr$ (e.g., \citealt[his Figure~2]{Bronfman2008});
3) a northern tangent near $25\degr$ (e.g., Table~1 of \citealt{Englmaier1999}, their ``inner Scutum tangent'').
The HMSFRs are not expected to follow the lines in Figure~\ref{lvdiagram} to any better than the 3--9~\kms\
velocity dispersion of the molecular cloud population (e.g., \citealt{Combes1991}, their Section~3.2.1).
Similarly, velocity dispersions measured from hydrogen profiles show characteristic broadening by $\sim7$~\kms,
that may be regarded as an upper limit due to the presence of blending features (e.g., \citealt{Burton1974}).

In the following, we present our results for each source and list the associated arms in the
last column of Table~\ref{tabinner}. Note that we also include two sources from previous measurements that are
of interest for a general discussion (G009.62$+$00.19 and G023.44$-$00.18).

\emph{G009.62$+$00.19}.
According to \citet{Sanna2009}, on the basis of a CO latitude-velocity analysis, this HMSFR
is associated with gas along the Norma Arm at a distance of $5.15^{+0.77}_{-0.66}$~kpc from the Sun, which
corresponds to a Galactocentric radius of 3.4~kpc for R$_0 = 8.38$~kpc. On the one hand, the near 3~kpc arm has
an LSR velocity blueshifted by $15$ to $20$~\kms\ at the same longitude.  The brightest CO emission from the near
3~kpc arm lies below the Galactic plane at Galactic longitudes greater than $\approx6\degr$ and is offset by about
$0.5\degr$ in latitude, with respect to the midplane of the Galaxy, at the Galactic longitude of the maser site
(Figure~\ref{bvdiagram}; see also Figure~3 of \citealt{Dame2008}). On the other hand, at $\ell \approx 9.5\degr$ the
Norma arm peaks at velocities between 0-10~\kms\ centered at about zero latitude, in agreement with the
Galactic location and LSR velocity of G009.62$+$00.19 (Figure~\ref{bvdiagram}).
We also note that the star-forming region appears to expand from the Galactic center at a velocity of
about 36~\kms, close to the Norma value of $\approx30$~\kms\ at zero longitude
\footnote{In Table~4, we have assumed as the peak LSR velocity
of the molecular cloud that from the CS(2--1) survey by \citet[+5~\kms]{Bronfman1996}, which is representative of the average velocity
over the whole region. In \citet{Sanna2009}, we assumed a slightly smaller velocity associated with the individual star-forming site
G009.62$+$00.19~E.} (Table~\ref{tabinner}).

\emph{G010.47$+$00.02}.
At a parallax distance of $8.55^{+0.63}_{-0.55}$~kpc from the Sun, our measurement translates
to a Galactocentric radius of 1.6~kpc, assuming R$_0 = 8.38$~kpc.
The LSR velocity of large-scale emission associated with G010.47$+$00.02 ($\sim 70$~\kms) falls in the
$\ell$-$v$ locus of the Connecting arm in the first quadrant, about 20~\kms\ lower
than a close by $\ell$-$v$ feature associated with the far 3~kpc arm (Figure~\ref{lvdiagram}).
At the longitude of the source, the far 3~kpc arm shows a bump of faint
emission below the Galactic plane (Figure~\ref{bvdiagram}; see also Figure~3 of \citealt{Dame2008}),
whereas G010.47$+$00.02 lies close to zero latitude, a few tens of parsec above the Galactic plane.
On the other hand, by comparing the velocity profile of the Connecting arm at $\ell \sim 10.5\degr$ with
the latitude-velocity maps of \citet[our Figure~\ref{bvdiagram}]{Bitran1997},
one can clearly see a strong CO feature centered at zero latitude at the LSR velocity of G010.47$+$00.02.
Therefore, we associate G010.47$+$00.02 with gas condensations belonging to the Connecting arm.

\emph{G010.62$-$00.38}.
The combined parallax measurement for G010.62$-$00.38 is $0.202 \pm 0.019$~mas, which corresponds to
a distance of $4.95^{+0.51}_{-0.43}$~kpc from the Sun and a Galactocentric radius of 3.6~kpc
(for R$_0 = 8.38$~kpc).
This prominent site of star formation has a combination of slightly negative LSR velocity ($-3$~\kms)
and negative Galactic latitude, which associates the star-forming region with a bright CO feature
belonging to the near 3~kpc arm (Figure~\ref{bvdiagram}, lower panel).
In this range of longitudes, the Norma and near 3~kpc arms are clearly separated by about 20~\kms\
at different distances below the Galactic plane (Figure~\ref{bvdiagram}).
G010.62$-$00.38 belongs to the W31 star-forming complex that has been long thought to have an intrinsic,
large, non-circular motion of several tens of \kms.
While hydrogen recombination lines from the large-scale \HII\ complex show velocities near zero,
which could be associated with either nearby or very distant gas, absorption features from different
molecules/transitions have been detected up to velocities of about 50~\kms\ (see discussion in, e.g.,
\citealt{Wilson1974,Caswell1975,Fish2003}). Along nearly circular orbits, the terminal velocity at
$\ell \approx 10\degr$ would be in excess of three times the absorption cutoff assuming a
IAU Galactic rotation speed, which would locate G010.62$-$00.38 on the near side of the tangent point
(e.g., \citealt{Fish2003}, their Figure~1).   While this argument has been questioned on the basis of
a general deficiency of CO and \HI\ gas within a Galactocentric radius of about 3~kpc
(e.g., Figure~6 of \citealt{Corbel2004}), the current parallax distance of 5~kpc from
the Sun indeed places W31 on the near edge of the Galactic gas hole, assuming the W31 cluster of \HII\ regions
is physically related (cf. \citealt{Corbel2004}, their Figure~8).
The observed absorption cutoff is due to the radial expansion of G010.62$-$00.38 from the Galactic center
($\sim 60$~\kms), as inferred from its full-space kinematics (Table~\ref{tabinner}).

\emph{G012.02$-$00.03}.
We measured a combined trigonometric parallax of $0.106 \pm 0.008$~mas for this source, corresponding to a distance of
$9.43^{+0.77}_{-0.66}$~kpc from the Sun and a Galactocentric radius of 2.1~kpc (for R$_0 = 8.38$~kpc).
In Figure~\ref{lvdiagram}, the $\ell$-$v$ position of G012.02$-$00.03 is associated with the locus of the
far 3~kpc arm in the first Galactic quadrant, which lies close to zero latitude at the
longitude of our source \citep[their Figure~3]{Dame2008}. For $\ell \approx 12\degr$, the current parallax
measurement locates the far 3~kpc arm at almost half the distance to the Galactic center of the near 3~kpc arm at
a similar longitudes (Figure~\ref{positions}), which has a direct implication for interpreting the nature
of the 3~kpc feature of our Galaxy (see Section~\ref{discus_dynamics}).

\emph{G023.44$-$00.18}.
\citet{Brunthaler2009} measured this massive star-forming region to be at a heliocentric distance of
$5.88^{+1.37}_{-0.93}$~kpc, or a Galactocentric radius of 3.8~kpc for R$_0 = 8.38$~kpc, in the general
direction of the northern Norma tangent in the first Galactic quadrant
(Figure~\ref{lvdiagram}). According to \citet[their Figure~10b]{Dame1986}, at the longitude of the source,
velocities in excess of $+100$~\kms\ are expected for the inner regions of the Norma arm, whereas
material in the Scutum arm would show LSR velocities which are more than 30~\kms\ below
that of G023.44$-$00.18. Therefore, we associate G023.44$-$00.18 as belonging to the Norma arm near
its tangent.

\emph{G023.70--00.19}.
The distance to this star formation site is $6.21^{+1.0}_{-0.80}$~kpc, which corresponds to a
Galactocentric radius of 3.7~kpc (for R$_0 = 8.38$~kpc).  This measurement locates G023.70$-$00.19 only
a few hundred pc away from G023.44$-$00.18 in a similar direction below the Galactic plane,
which argues for an association with the Norma arm. As shown in Figure~\ref{lvdiagram}, its LSR
velocity of about $+77$~\kms\ associates the star-forming region to gas emission at the low-longitude
edge of a ``gas hole'' near $\ell \sim 25\degr$. According to \citet[their Figure~2]{Cohen1980},
molecular gas at velocities between 70-90~\kms\ and Galactic longitudes between $23\degr$ and $25\degr$
is associated with the Norma arm. At a similar longitude, gas belonging to the nearby Scutum arm
shows LSR velocities less than $+60$~\kms\ that involves CO emission below the gas hole.
Therefore, we associate G023.70--00.19 with gas condensations in the Norma arm.

\section{Discussion}\label{discuss}

Together with the two previous measurements listed in Table~\ref{tabinner}, the HMSFRs presented here
sample the inner regions of the Milky Way at Galactic longitudes from $+9.6\degr$ to $+23.7\degr$ and
Galactocentric radii between~1.6 and~3.8~kpc.  In the inner Galaxy, the mass distribution
of the Galactic (nested) bar(s) provides non-axisymmetric gravitational perturbations on gas
and stellar orbits, whereas a nearly flat rotation curve is generally measured between
Galactocentric radii of about 4 to 13~kpc (e.g., \citealt{Reid2009b}, \citealt{Honma2012}).
Under the influence of a bar potential, highly non-circular streamlines of gas and stellar orbits
are indeed expected as for the families of periodic orbits, $x_1$, $x_2$, and $x_3$
(e.g., \citealt{Contopoulos1980,vanAlbada1982,Athanassoula1992a}); these streamlines would
appear on $\ell-v$ diagrams as parallelogram-shaped features (e.g., \citealt{Binney1991,Bureau1999}).
A comparison of the measured LSR velocities with those expected for circular orbits
(e.g., \citealt{Englmaier1999}) at the longitudes and Galactocentric distances of our sample shows,
in general, differences of several tens of \kms\ (e.g., \citealt{Sanna2009}).
From Table~\ref{tabinner}, a weighted average of the rotation speed of HMSFRs in the inner Milky Way differs
by about $-40\pm5$~\kms\ from the rotation speed at the Solar circle of
$\Theta_0=243$~\kms\ \citep{Reid2013}. This value is significantly lower than the average peculiar motion derived
for HMSFRs at Galactocentric radii greater than 4~kpc, which lag Galactic rotation by less than 10~\kms\
(e.g., \citealt{Xu2013}). We note that this result does not depend sensitively on the value of $\Theta_0$ assumed
in the calculation. Evidence for a substantially flat rotation curve which drops inwards at small Galactocentric radii
are commonly observed for barred galaxies with rotation velocities comparable to the Milky Way (e.g., \citealt{Sofue1999}).

Moving from the star-forming region closest to the Galactic center (G010.47$+$00.02) to the one farthest from
the center (G023.44$-$00.18), we start to fix the positions of the major arm-like features of the inner Milky Way
on a coherent face-on view (Figure~\ref{positions}). In the following, we describe these constraints in brief.
Within about 3~kpc of the Galactic center and outside the CMZ, three CO features pointed out on the $\ell$-$v$
diagram of Figure~\ref{lvdiagram} dominate the gas kinematics, the near and far 3~kpc arms (yellow lines) and the Connecting
arm (green line). The Connecting arm, as fixed by the position of G010.47$+$00.02 at R$=1.6^{+0.2}_{-0.1}$~kpc, apparently runs
along the far edge of the long bar, probably closer to the Galactic center than the far 3~kpc arm, constrained at a similar
longitude by the position of G012.02$-$00.03 at R$=2.1^{+0.5}_{-0.3}$~kpc. The distance measurement to G010.62$-$00.38 further
constrains the position of the near side of the 3~kpc arm at a Galactocentric radius (R$=3.6^{+0.4}_{-0.4}$~kpc) almost twice
that of G012.02$-$00.03. Despite early claims that the near 3~kpc arm was deficient in star formation activity
(e.g., \citealt{Lockman1980}), at the position of our sources the CO velocity-integrated maps show local peaks of emission (Figure~3
of \citealt{Dame2008}), which pinpoint enhanced star formation such as in the prominent W31 complex.  This is also in
agreement with the recent detection of several, massive, star formation sites from the Methanol Multibeam Survey
that are associated in longitude, latitude, and velocity with the brightest CO emission along the 3~kpc arms
(cf. Figure~2 of \citealt{Green2009} and Figure~3 of \citealt{Dame2008}).

In the first Galactic quadrant, outward of 3~kpc from the Galactic center but within the GMR,
an annulus of less than 2~kpc in radius encompasses the northern Scutum-Centaurus, Norma, and near
3~kpc arm tangents (Figure~\ref{positions}).
This is evident, for instance, in the Galactic Ring Survey of extended $^{13}$CO emission along
the Galactic plane, with prominent concentrations at $\ell \sim 31\degr$ and $23\degr$
\citep[their Figure~1]{Jackson2006}.
While interarm distances are expected to decrease as the arms converge to the end of
the (long) bar at a Galactocentric azimuth near $45\degr$ (\ie\ the angle between the Sun and the source
as viewed from the Galactic center, $\beta$), our distance measurements for G009.62$+$00.19
and G010.62$-$00.38 show that the Norma and near 3~kpc arms nearly overlap at a Galactic longitude
of $10\degr$ ($\beta \sim 14\degr$) on a face-on view of the Milky Way (Figure~\ref{positions}).
On the other hand, a $b$-$v$ analysis of the CO emission from the arms at this longitude reveals
that their gas distributions do not mix, with the near 3~kpc arm passing under (at lower Galactic
latitudes) the Norma arm (Figure~\ref{bvdiagram}).

\subsection{Topics on the Inner Arm Features}\label{discus_dynamics}

The small number of measured distances to sources in the inner portion of the Galaxy precludes
a thorough discussion on individual arm features at this time.
Similarly, an analysis of the inner rotation curve (R$<4$~kpc), which one would expect to
have a strong azimuthal dependence, is also premature.
However, our measurements still allow us to start discussing two specific topics.

Firstly, if the star-forming region G010.47$+$00.02 is located in the Connecting
arm at a longitude of $\ell \sim 10.5\degr$, then it is close to the far edge of both the
Galactic and ``long'' bars in the first Galactic quadrant (Figure~\ref{positions}).
According to \citet[their Section~4]{Roberts1979}, streamlines of gas crossing the major axis
of a bar strongly decelerate near their maximum radial excursion, where gas approaches a potential
minimum (e.g., Figure~5 and~6 of \citealt{Roberts1979}).
A major effect of slowing down abruptly would be that gas piles up until a shock occurs
(i.e., the locus of the Connecting arm). We find for G010.47$+$00.02 that the V$_s^{\rm G.C.}$
velocity component of $122\pm16$~\kms\ (in the direction of Galactic rotation) is slower by almost a factor of 2 than
the circular rotation speed beyond 4 kpc in the Galaxy. While the rotation curve is expected to drop
toward the center owing to less mass enclosed, a nearby HMSFR belonging to the far 3~kpc arm (G012.02$-$00.03)
shows a rotation speed of $215\pm27$~\kms. This evidence suggests that the V$_s^{\rm G.C.}$ value of
G010.47$+$00.02 is likely related to the local gas dynamics, and represents an argument in favor of the shock
phenomenon as the origin of its peculiar motion. The position of the Connecting arm with respect to the axis
of the bar(s), and the direction of the gas flow along the Connecting arm
itself (U$_s^{\rm G.C.}$ component in Table~\ref{tabinner}), should be intimately related to the
presence of an inner Lindblad resonance (ILR) close to the Galactic center
(e.g., \citealt{Roberts1979,Athanassoula1992,Athanassoula1999}).
If an ILR is present, or equivalently if the $x_2$ family of stable periodic orbits exist
(i.e., orbits oriented perpendicular to the bar major axis), the shock locus should be
offset toward the leading edge of the bar; the larger the $x_2$ orbit
along the bar minor axis, the bigger will be the offset of the Connecting arm with respect to the
major axis of the bar (e.g., Figure~5 of \citealt{Athanassoula1992}). The position of
G010.47$+$00.02 is offset by about 1.3~kpc (i.e., $\approx 2\sigma$) from the direction of the major
axis of the long bar. While we cannot rule out an overestimate of our distance measurement by $2\sigma$,
the current measurement supports the Connecting arm being shifted forward (clockwise
in phase) toward the far side of the bar in the first Galactic quadrant, or alternatively it
supports the presence of an ILR.  \citet[their Figure~6]{Roberts1979} also predict that a
shock-focusing phenomenon should develop where an ILR is present.
Given that, forced by the ILR, the Connecting arm would be inclined with respect to the orientation
of the bar's major axis, then streamlines of gas at different Galactocentric radii would enter the shock
front at different angles.
As a result, gas along the Connecting arm in the outer bar regions would have a net inward motion toward the
Galactic center with velocities of some tens of \kms; at some Galactocentric
radii inwards, the gas motion along the shocked layer would invert direction, showing an instantaneous
velocity component outward from the Galactic center with velocities
of the order of 100~\kms.  The overall effect of the shock-focusing
phenomenon would be to channel gas at some Galactocentric radii and enhanced star formation activity
would be expected in the region of converging flows. According to \citet{Roberts1979}, a detection of an
outward motion along the Connecting arm would be an independent confirmation for the presence of an
ILR in the bar region.
Our data show a combination of offset shock and inward motion that would confine the shock-focusing
phenomenon (if present) to Galactocentric radii within about 1.6~kpc from the Galactic center.

A second topic concerns the nature of the near and far 3~kpc arms (e.g., see the review in
Section~3.2 of \citealt{Green2011}).  We follow the interpretation of the 3~kpc arms as a
continuous physical structure as opposed to, for instance, a combination of two,
separate, lateral arms.  This assumption is based on a better match of the former
interpretation to the $\ell$-$v$ locus of the 3~kpc arm features
(cf. Figure~7, 10, and~11 in \citealt{Green2011}). In this context, two main ideas have been
forwarded: 1) that of a circular ring with an expanding motion away from the Galactic center
(e.g., \citealt{vanderKruit1971}), and 2) that of an elliptical streamline of gas around the
bar region (e.g., \citealt{Peters1975}).  An important difference between the two models is related
to the position of the 3~kpc arm with respect to the radius of the corotation resonance
($r_{\rm CR}$, associated with the rotation pattern of the bars): elliptical orbits are expected
inside $r_{\rm CR}$, whereas nearly circular orbits should occur outside (e.g., \citealt{Contopoulos1980}).
The Galactocentric radius of the near 3~kpc arm at the position of G010.62$-$00.38 (R$=3.6^{+0.4}_{-0.4}$~kpc)
is almost two times greater than the Galactocentric radius of the far 3~kpc arm, as inferred from
the position of G012.02$-$00.03 (R$=2.1^{+0.5}_{-0.3}$~kpc).  This does not match the expected
constant radius for a circular ring and seems to favor a highly eccentric orbit.
If the interpretation of elliptical orbits holds, we also note that the maximum Galactocentric radius
measured for the  3~kpc arms (for G010.62$-$00.38) provides a lower limit to the radius of
corotation at $r_{\rm CR} \gtrsim 3.6$~kpc, consistent with previous findings from the literature
(e.g., \citealt{Englmaier1999}, their Section~4.5).

In Figure~\ref{positions}, the elliptical model by \citet{Green2011} for the 3~kpc feature provides a good match to
the position of the two HMSFRs. This model assumes that gas belonging to the 3~kpc arms flows along an
elliptical streamline and has a constant angular momentum at any given point of the ellipse, with a value of 320~km\,s$^{-1}$\,kpc.
Locally, the tangential velocity is given by the ratio of the angular momentum
and the Galactocentric radius at that point. Our measured proper motions for G010.62$-$00.38 and G012.02$-$00.03
give values of the angular momentum higher than 400~km\,s$^{-1}$\,kpc. By increasing the angular momentum value, the
model by Green et al. still provides a reasonable fit to the data, but requires a shift of the ellipse orientation
at smaller angles with respect to the Sun-Galactic center direction (e.g., Figure 8 of
\citealt{Green2011}).
Whether or not the orientation of the ellipse fits better the direction of the long ($\beta \sim 45\degr$)
or the Galactic bar ($\beta \sim 20\degr$) would be a hint as to which bar component dominates the gas response.
While our data still do not constrain the ellipse orientation with sufficient accuracy, we should be able to assess
these issues with stronger statistical support when more parallax measurements in the inner Milky Way will be available.

\acknowledgments

This work was partially funded by the ERC Advanced Investigator Grant GLOSTAR (247078).
This work made use of the Swinburne University of Technology software
correlator, developed as part of the Australian Major National Research
Facilities Programme and operated under licence.

{\it Facilities:} \facility{VLBA}.



\appendix

\section{Parallax and Proper Motion Fitting Details}\label{appendix}

Maser spots for parallax fitting were selected according to the following criteria: 1) spots persisting
over one year that belong to isolated features, in order to avoid emission blended between different maser
centers; 2) compact maser spots, unresolved by the VLBA beam or slightly resolved but with a stable,
deconvolved, position angle; 3) strong maser spots ($\sim 1-10$~Jy~beam$^{-1}$) with typical signal-to-noise
ratios of more than a hundred.

\emph{G010.62$-$00.38}.
For the purposes of a maser reference position, we employed three spots at the LSR velocities of $-14.2$,
$-1.1$, and +1.0~\kms\ from the H$_2$O maser distribution measured within the field of view of the VLBA.
Among the calibrators observed in K band, the two stronger QSOs in Table~\ref{tabsou}, J1821$-$2110 and J1751$-$1950,
served as a background reference position and their images from the first epoch data are shown in Figure~\ref{calib}.
The angular separation on the plane of the sky ($\theta_{\rm sep}$) between the background QSOs and the target
maser is reported in Table~\ref{tabsou}.
The ICRF J1751$-$1950 was detected at each epoch above a 5\,$\sigma$ level; at the third epoch, the peak intensity
of J1821$-$2110 fell below a 3\,$\sigma$ level due to poor phase stability between the maser and the calibrator
(because of poor weather conditions) and was not used. Due to the lack of the third epoch data, parallax fitting with
the calibrator J1821$-$2110 results in a higher correlation between the parallax sinusoid and the (linear) proper
motion component from each maser spot. Since we expect no detectable proper motion for the extragalactic sources,
parallax fitting with J1821$-$2110 was constrained with proper motions  determined from J1751$-$1950 for each spot.
The error-floors determined by combining the measurements of the three maser spots with respect to the
two QSOs were $\pm 0.05$~mas in the E--W direction and $\pm 0.07$~mas in the N--S direction.
Since water maser cloudlets have typical proper motions of tens of \kms, determining a secular proper motion for the
HMSFR requires to correct for maser velocity components. We estimated this contribution from the average proper motion
of all maser features (80) determined as  in \citet{Sanna2010a} with respect to the reference spot at +1.1~\kms,
used for the parallax fitting and with the more accurate proper motion measurement.
The average, internal, velocity components are $-0.01 \pm 0.03 $~mas~yr$^{-1}$ toward the east and
$+0.28 \pm 0.02 $~mas~yr$^{-1}$ toward the north, where we report the standard error of the mean.
Thus, the secular proper motion of the HMSFR is estimated to be $-0.366 \pm 0.081 $~mas~yr$^{-1}$ and
$-0.600 \pm 0.055 $~mas~yr$^{-1}$ in the east and north directions, respectively.
We explicitly note that these small uncertainties may still be affected by a further uncertainty of several tenth of
mas~yr$^{-1}$, due to the complexity of \Gm\ as a cluster of young stellar objects (e.g., \citealt{Liu2011}).
At our measured distance, these values correspond to $-8.6 \pm 1.9$~\kms\ and $-14.1 \pm 1.3$~\kms\ eastward and northward,
respectively. Completing the kinematic information, we assume an LSR velocity of $-3.0 \pm 2.7 $~\kms\ for the HMSFR \Gm,
obtained from the large-scale rest velocity of the CS\,(1$-$0) and NH$_3$\,(1,1) line emission \citep{Anglada1996}.

\emph{G010.47$+$00.02}.
As a maser reference position, we combined the positions of seven spots associated with seven
distinct maser cloudlets spread over a region of about 0.4~arcsec in size (Table~\ref{tabres}).
Imaging of the calibrators for G010.47$+$00.02 was optimized by setting an elevation cutoff of $25\degr$ for each antenna.
Among the set of calibrators observed in combination with G010.47$+$00.02, only the ICRF calibrator J1751$-$1950,
at an angular offset of $3\degr$ from the target maser (Table~\ref{tabsou}), had a distinct peak of emission at all
epochs and has been used as a background reference position in the parallax measurement (Figure~\ref{calib}).
The error-floors determined from the simultaneous fitting of these seven maser spots were $\pm 0.01$~mas and
$\pm 0.09$~mas toward the E--W and N--S directions, respectively.
As for G010.62$-$00.38, we estimated the secular proper motion of the star-forming region by subtracting the
relative motion of all measured maser features (11) with respect to the reference spot at +68.0~\kms\ (Table~\ref{tabres}).
The average internal velocity components are $+0.073 \pm 0.006 $~mas~yr$^{-1}$ toward the east and
$-0.22 \pm 0.08 $~mas~yr$^{-1}$ toward the north, where we report the standard error of the mean.
Thus, the total motion of the whole source is estimated to be $-3.860 \pm 0.015 $~mas~yr$^{-1}$ and
$-6.403 \pm 0.076 $~mas~yr$^{-1}$ in the east and north directions, respectively.
At our measured distance, these values correspond to $-156.4 \pm 0.6$~\kms\ and $-259.5 \pm 3.1$~\kms\
eastward and northward, respectively.

\emph{G012.02$-$00.03}.
The 12.2~GHz methanol maser from G012.02$-$00.03 consists of a single spectral feature
persisting during the four observing epochs with emission extended over an area of a few mas (squared). The brightness
distribution is centrally peaked and we employed the peak positions of two velocity
channels, at +108.0 and +108.8~\kms, as maser reference positions. While the U-band calibrators were easily detected at
all epochs, only the closer QSO J1808--1822 (see Table~\ref{tabsou}) had a stable spatial morphology (deconvolved size)
suitable for an accurate parallax measurement.
The error-floors from these combined measurements were $\pm 0.01$~mas and $\pm 0.19$~mas toward the E--W and N--S directions,
respectively. Methanol masers move with typical velocities of only a few \kms\
that are very close to the systemic velocity of the molecular cloud core they are associated with.
Therefore, we will neglect their contribution for computing the secular proper motion of \Gq.
By averaging the proper motion inferred from the two maser spots, we obtain Galactic velocity components of
$-183.6 \pm 1.0$~\kms\ and $-346.7 \pm 11.8$~\kms\ eastward and northward, respectively.

\begin{figure*}
\centering
\includegraphics[angle= 0, scale= 0.9]{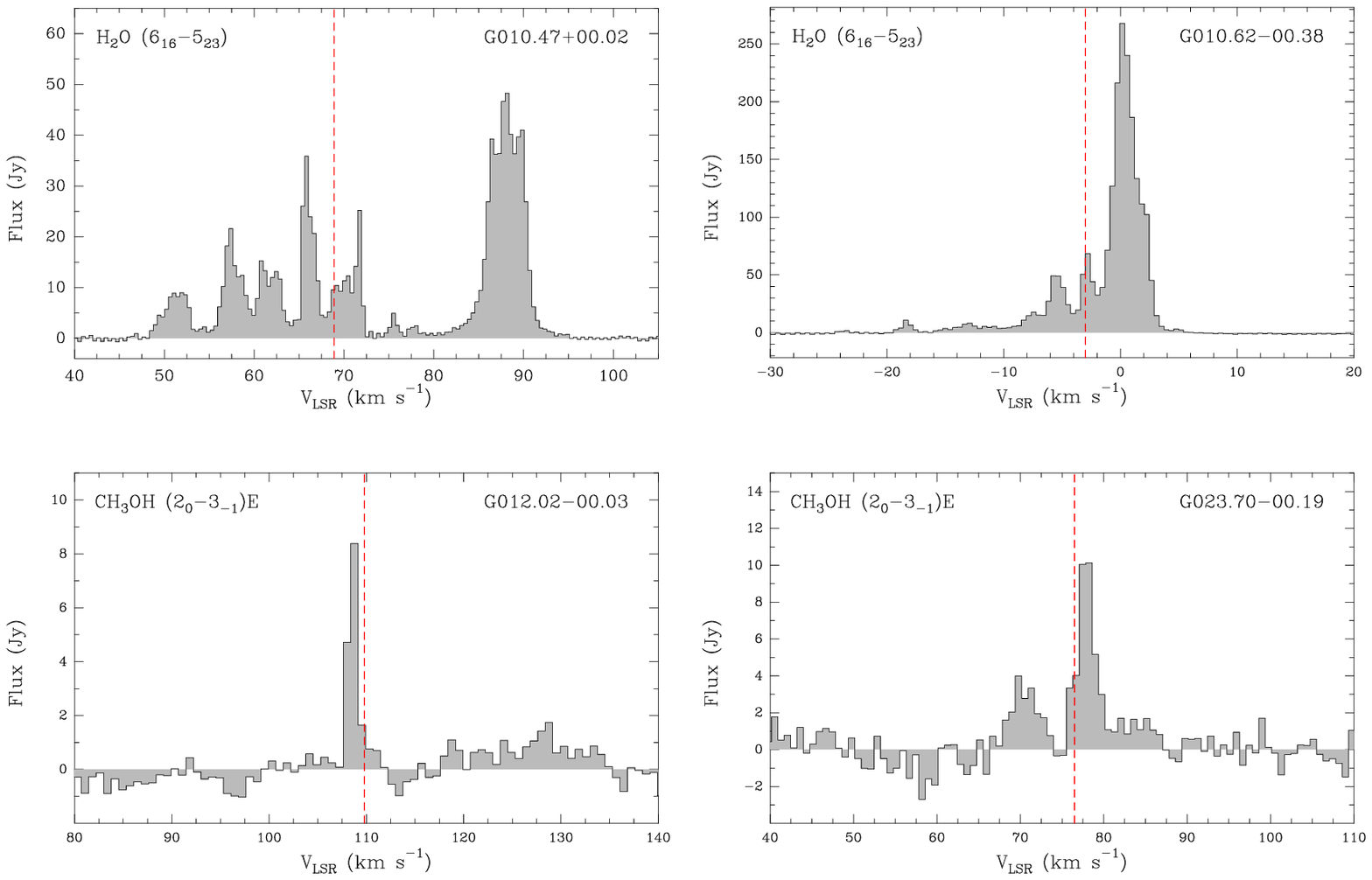}
\caption{Total-power (Stokes~I) spectra of the maser emission from the first epoch data for the inner Galaxy sources.
These profiles were obtained by averaging the total-power spectra of all VLBA antennas with the task POSSM of AIPS and
removing the baseline within the CLASS package of GILDAS (http://www.iram.fr/IRAMFR/GILDAS). Each dashed vertical line
marks the systemic velocity (V$_{sys}$) of the large-scale molecular cloud hosting the star-forming region (see Table~\ref{tabinner}
and Appendix~\ref{appendix}).
See the electronic edition of the Journal for a color version of this figure.
\label{spectrum}}
\end{figure*}

\begin{figure*}
\centering
\includegraphics[angle=0,scale=.6]{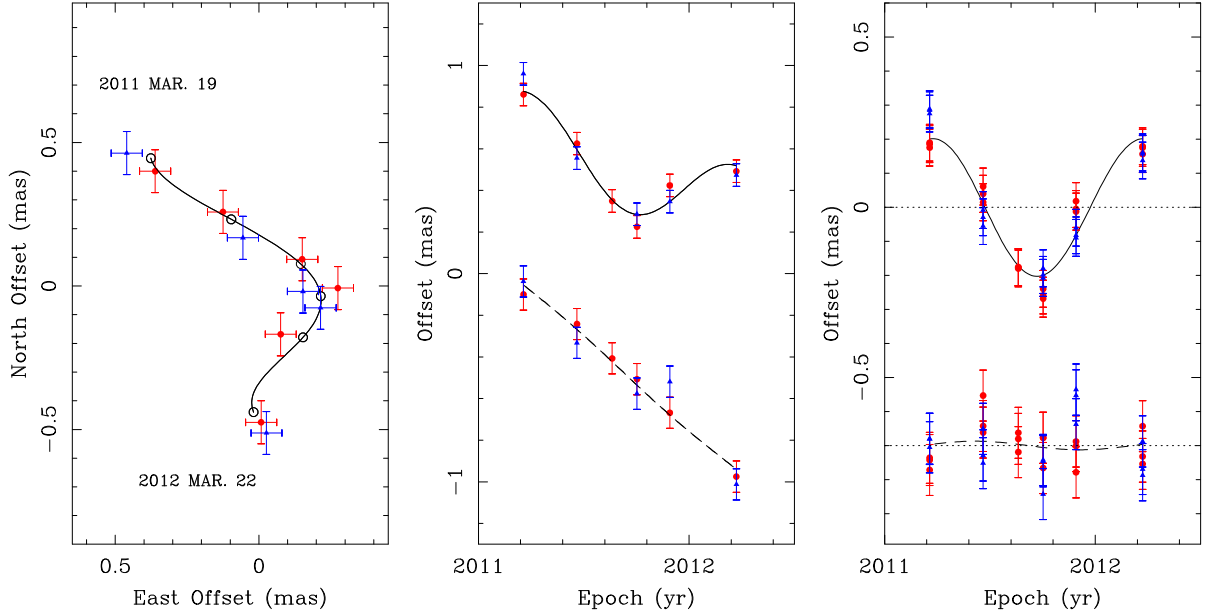}
\caption{Results of the ``combined'' parallax fit for G010.62$-$00.38. \textit{Left Panel:} Sky projected motion of a
single maser spot at +1.0~\kms\  with respect to J1751$-$1950 (red circles) and J1821$-$2110 (blue triangles)
with first and last epochs labeled. The empty circles and the line show the best-fit position
offsets and the trajectory, respectively. \textit{Middle Panel:} The position offsets of the maser at +1.0~\kms\
along the east and north directions versus time. The best-fit model in east and north directions are shown as continuous
and dashed lines, respectively. \textit{Right Panel:} Same as the middle panel but with fitted proper motions subtracted
(i.e. parallax curve) and all maser spots measurements drawn for each QSO (3 maser spots and 2 QSOs). The north offset
data have been shifted for clarity. See the electronic edition of the Journal for a color version of this figure. \label{paralg10}}

\end{figure*}

\begin{figure*}
\centering
\includegraphics[angle=0,scale=0.6]{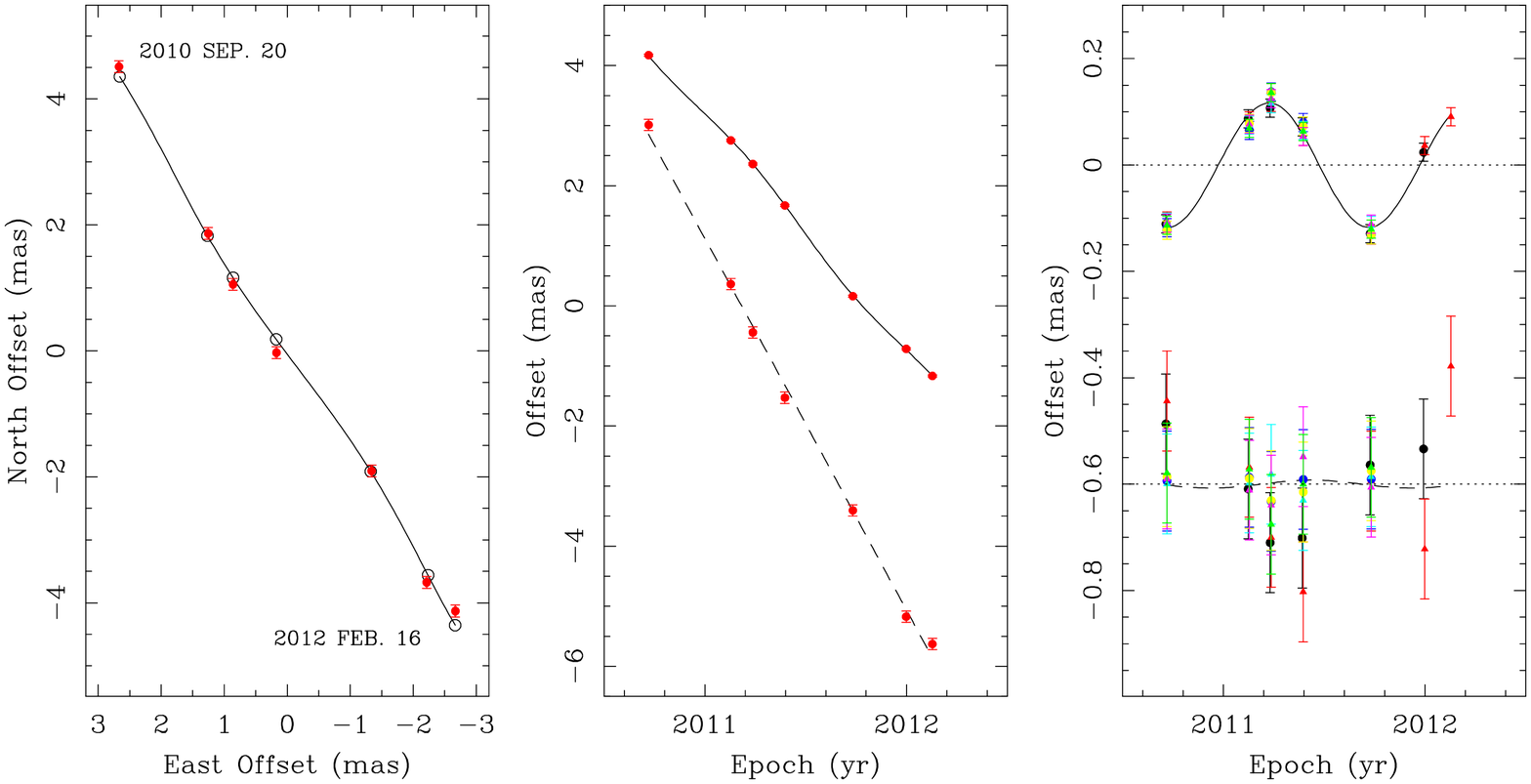}
\caption{Results of the ``combined''  parallax fit for G010.47$+$00.02. Similar to Figure~\ref{paralg10} with positions for a maser spot at
68.0~\kms\ (red circles), which lasted for the higher number of epochs,  measured relative to J1751$-$1950 (left and middle panels). For the
parallax curve (right panel), all maser spots are drawn with different colors (7 maser spots and 1 QSO).
See the electronic edition of the Journal for a color version of this figure.\label{paralg1047}}
\end{figure*}

\begin{figure*}
\centering
\includegraphics[angle=0,scale=0.6]{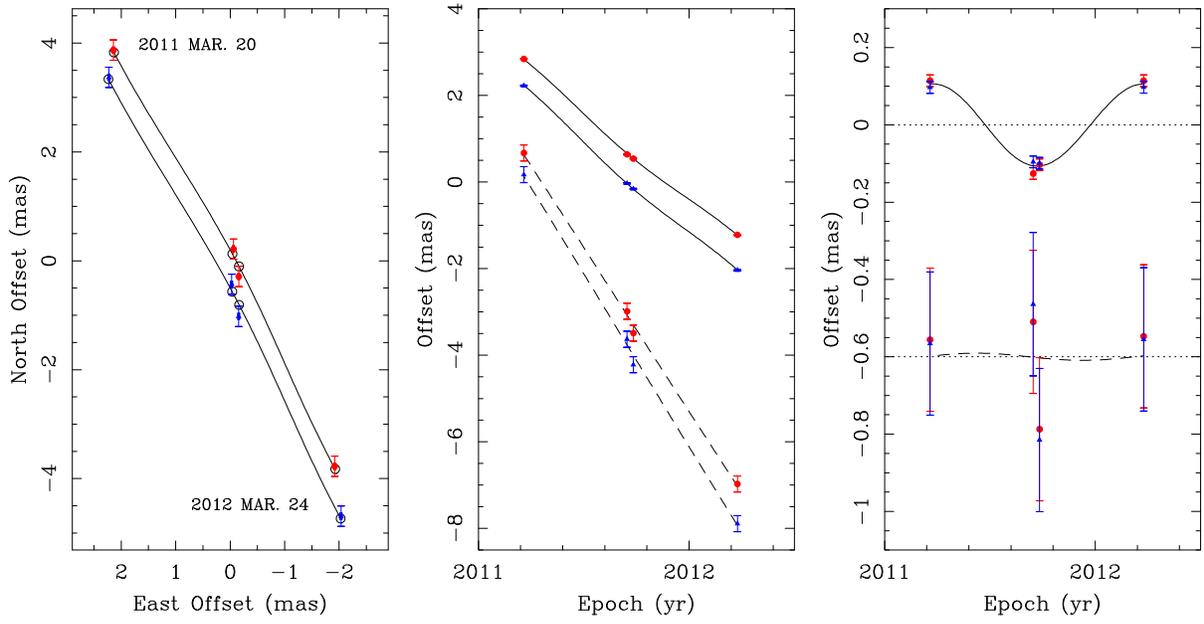}
\caption{Results of the ``combined''  parallax fit for G012.02$-$00.03. Similar to Figure~\ref{paralg10} with maser positions for spots at
+108.0 (blue triangles) and +108.8~\kms\ (red circles) measured relative to J1808$-$1822 (2 maser spots and 1 QSO).
See the electronic edition of the Journal for a color version of this figure. \label{paralg12}}
\end{figure*}

\begin{figure*}
\centering
\includegraphics[angle=0,scale=0.6]{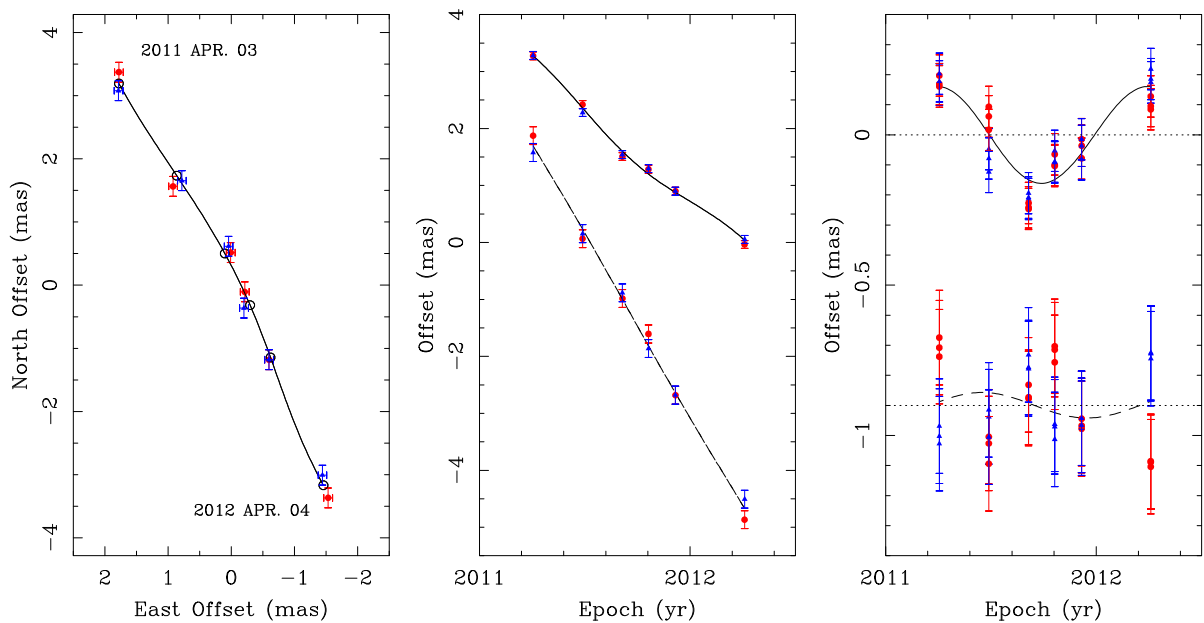}
\caption{Results of the ``combined''  parallax fit for G023.70$-$00.19. Similar to Figure~\ref{paralg10} with maser positions for spot at
+79.0~\kms\ with respect to J1825$-$0737 (red circles) and J1846$-$0651 (blue triangles), and all maser spots drawn for the parallax curve
in the right panel (3 maser spots and 2 QSOs).
See the electronic edition of the Journal for a color version of this figure. \label{paralg23}}
\end{figure*}

\begin{figure*}
\centering
\includegraphics[angle=0,scale=0.9]{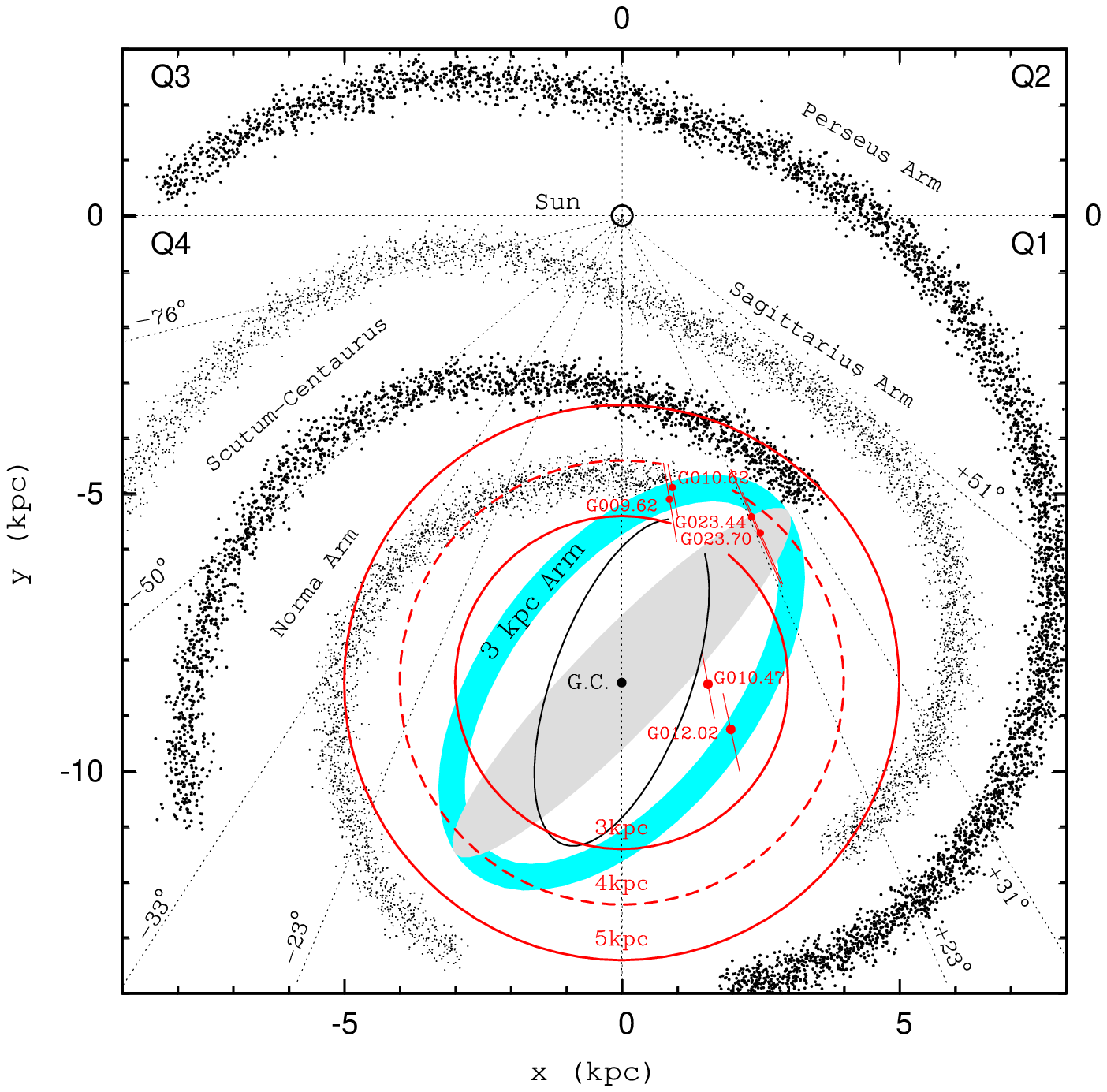}
\caption{Schematic view of the spiral arms of the Milky Way across the four Galactic quadrants after
\citet{Taylor1993} with updates. The best-value of R$_0 = 8.38$~kpc from  \citet{Reid2013} is assumed.
The location of the Galactic bar (i.e., the bulge) and the central ``long'' bar as from \citet{Churchwell2009}
are also reported as an open and light-grey ellipsoids, respectively. Positions of the inner Milky Way sources
from Table~\ref{tabinner} are labeled as well (red dots) together with error bars. Dotted lines drawn from the Sun position
across the first and fourth Galactic quadrants mark the tangents to major spiral arms as from \citet{Vallee2008}.
The near 3~kpc arm tangents at $\pm 23\degr$ are adopted following \citet{Dame2008}.
The azure ellipsoid outlines the locus of the 3~kpc arms following the best fit model to the
$\ell$-$v$ position of 6.7~GHz masers by \citet{Green2011}, with a total width of 0.5~kpc.
See the electronic edition of the Journal for a color version of this figure. \label{positions}}
\end{figure*}

\begin{figure*}
\centering
\includegraphics[angle=0,scale=0.9]{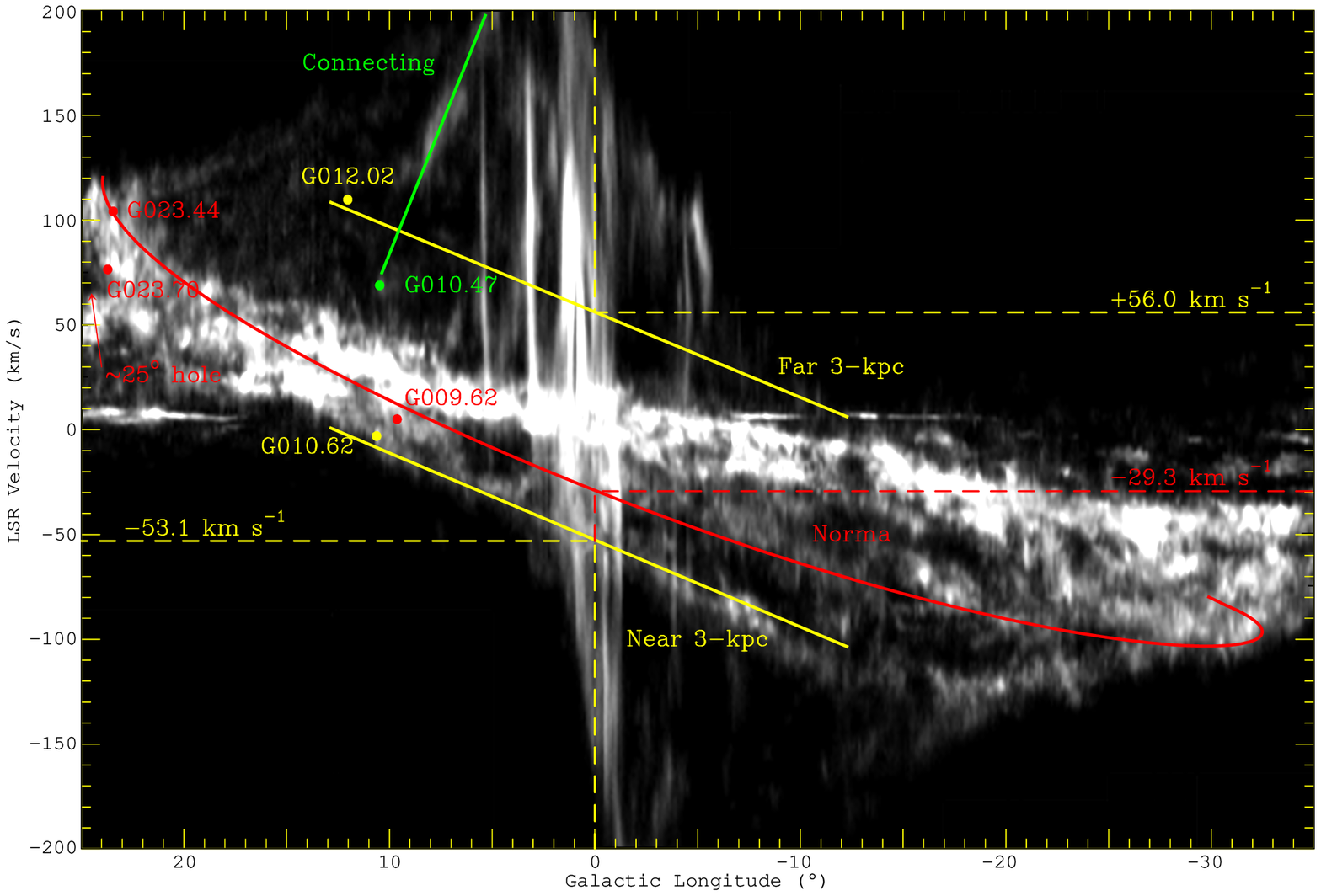}
\caption{Longitude-velocity diagram of the CO emission integrated within $-1\degr \leq b \leq +1\degr$ from
\citet{Dame2001}. The gray scale is CO intensity ranging from~0 (black) to 4~K~arcdeg (white).
Lines draw the $\ell$-$v$ positions of gas associated with the inner arms: the yellow lines
mark the near and far 3~kpc arms from \citet{Dame2008}, the green line marks the Connecting arm
(e.g., \citealt{Fux1999}, his Figure~1) and the red line marks a best-guess for the Norma arm locus
(see Section~\ref{association}). The expanding velocity components of the
Norma and 3~kpc arms at $\ell = 0\degr$ are marked for each fit as well. Positions of star-forming regions listed in
Table~\ref{tabinner} are plotted and color-coded according to the associated arms, together with the position
of the ``gas hole'' near $\ell \sim 25\degr$ separating the Norma and Scutum-Centaurus arms \citep[their Figure~2]{Cohen1980}.
See the electronic edition of the Journal for a color version of this figure.\label{lvdiagram}}
\end{figure*}

\begin{figure*}
\centering
\includegraphics[angle=0,scale=1.5]{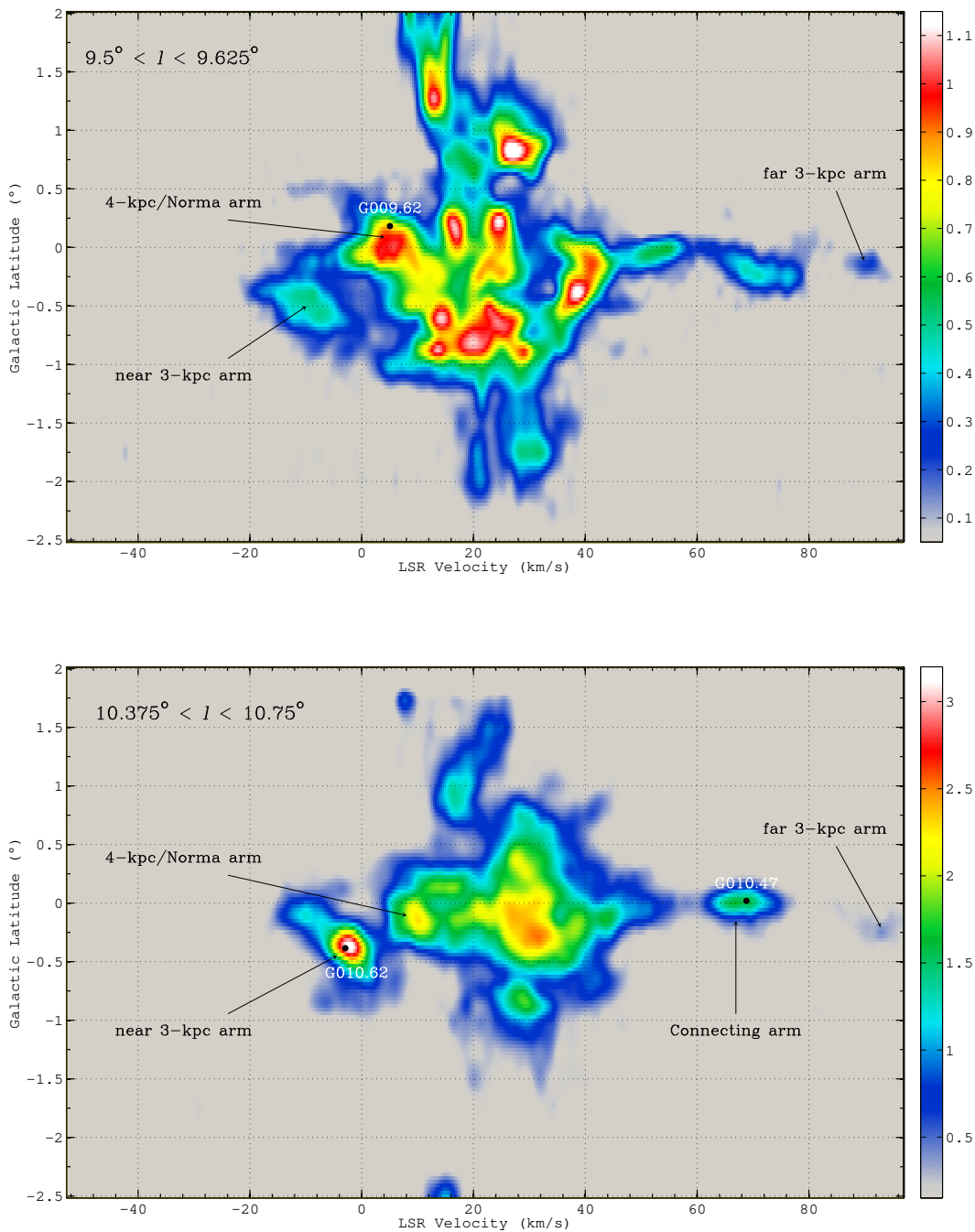}
\caption{Latitude-velocity diagrams of the CO emission integrated within the range of longitudes where the Norma and near 3~kpc arms
intersect in Figure~\ref{positions}. CO data are from the 1.2~m Cerro Tololo survey of \citet{Bitran1997}. The CO emission
was integrated within the ranges of Galactic longitudes indicated in the upper left corner of each panel.
The color scale on the right-hand side of each panel corresponds to the CO intensity range in K~arcdeg.
Positions of prominent arms in the inner Milky Way are labeled in each panel together with the associated sources
according to Table~\ref{tabinner}.
See the electronic edition of the Journal for a color version of this figure. \label{bvdiagram}}
\end{figure*}

\begin{figure*}
\centering
\includegraphics[angle= 0, scale= 1.0]{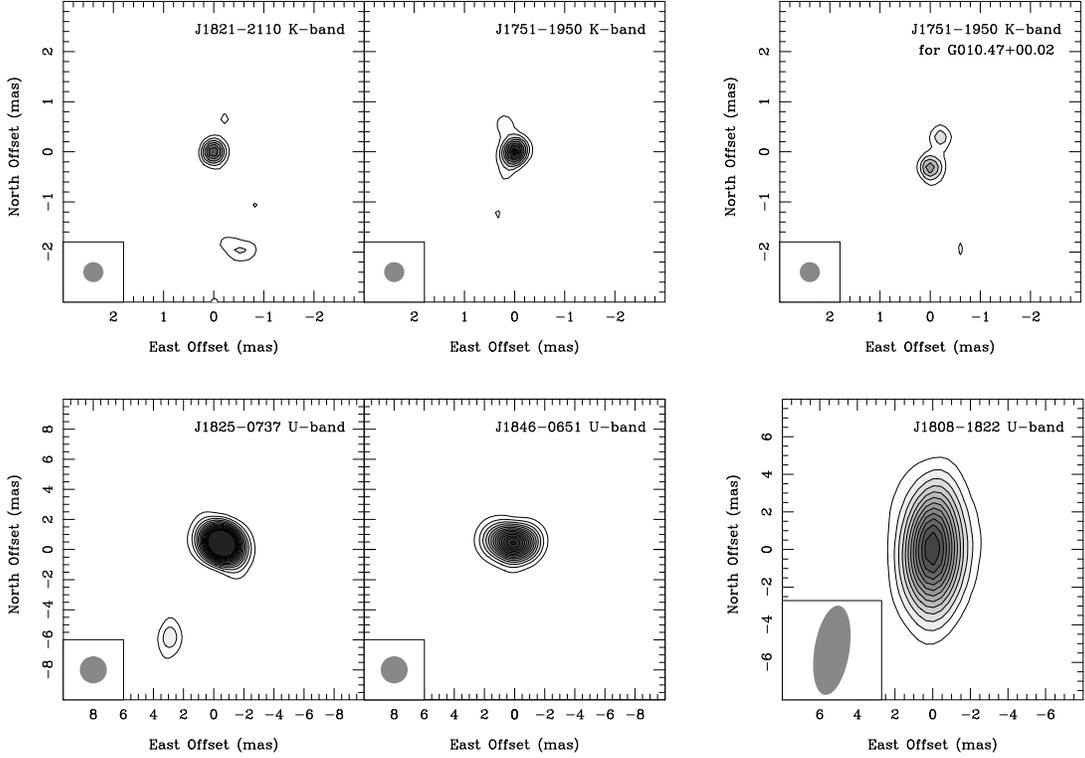}
\caption{Images of the background continuum sources in K and U bands near the target masers used
for parallax purposes. All images are from the first epoch observations (see Table~\ref{tabobs}).
Source names are indicated in the upper right corner and restoring beams are drawn in the lower left
corner of each panel. Contour levels start at $3\,\sigma$ by steps of $2\,\sigma$ and $3\,\sigma$ for the K and U-band calibrators,
respectively (see Table~\ref{tabsou}).  K-band calibrators for G010.62$-$00.38 and G010.47$+$00.02 were imaged with an HPBW
of 0.4~mas (round), U-band calibrators for G023.70$-$00.19 were imaged with an HPBW of 1.8~mas (round), and the
calibrator for G012.02$-$00.03 was imaged with a natural restoring beam.  \label{calib}}
\end{figure*}

\emph{G023.70$-$0.19}.
For the parallax measurement of G023.70--00.19 we made use of three maser spots, two of them
are associated with the peak position of the 12.2~GHz methanol maser emission (at $+77.5$ and $+79.0$~\kms) and a third spot
is associated with an isolated maser cloudlets at about 21~mas to the south of the peak emission (at $+76.7$~\kms).
Only measurements with the two closer calibrators (see Table~\ref{tabsou}), J1825$-$0737 and J1846$-$0651,
gave individual error-floors in the E-W direction smaller than 0.1~mas for each maser spot and were used in the
parallax estimate. These final error-floors were $\pm 0.06$~mas and $\pm 0.15$~mas toward the E--W and N--S directions,
respectively.
By averaging the proper motion inferred from the three maser spots, we obtain Galactic velocity components of
$-94.4 \pm 2.0$~\kms\ and $-187.4 \pm 4.0$~\kms\ eastward and northward, respectively.
Toward the maser position, CS\,(2$-$1) line emission shows an LSR velocity of $+68.3$~\kms\ (\citealt{Bronfman1996}), about
ten \kms\ slower than the peak velocity of the $\rm H110\alpha$ line at the same position \citep[$+76.5$~\kms]{Sewilo2004}. In order to
resolve the near/far kinematic distance ambiguity, Sewilo et al. observed the H$_2$CO\,($1_{10}-1_{11}$) line in absorption
against the \HII\ region up to velocities of $+86$~\kms, with an H$_2$CO absorption component at the same velocity of the CS
line. Following these evidence, we assume a central LSR velocity for the HMSFR G023.70--00.19 of $+76.5 \pm 10$~\kms, that better matches
the 12~GHz CH$_3$OH maser velocities of Figure~\ref{spectrum}, which usually trace more quiescent gas dynamics close to the systemic
velocity of the region ($< 5-10$~\kms).






\clearpage

\clearpage

\clearpage

\clearpage


\begin{deluxetable}{cllcll}
\tabletypesize{\scriptsize}
\tablecaption{Observation Information\label{tabobs}}
\tablewidth{0pt}
\tablehead{
           & \colhead{Exp. code:} & \colhead{BR145M} & \colhead{BR145O} & \colhead{BR145Q} & \colhead{BR145U} \\}

\startdata

Target     &                      & G010.62$-$00.38    & G023.70$-$00.19    & G012.02$-$00.03 & G010.47$+$00.02       \\
Maser line & & 22.2 GHz H$_2$O  & 12.2 GHz CH$_3$OH  & 12.2 GHz CH$_3$OH  &  22.2 GHz H$_2$O  \\
V$\rm _{LSR}$ (km s$^{-1}$) & & 1.0 & 79.0 & 108.0 & 65.0 \\
prep.survey & & BR145AL & BR145AK & BR145AH & BR145AW\\
Epoch: & & & & & \\
1 & & 2011 MAR 19  & 2011 APR 03  & 2011 MAR 20  & 2010 SEP 20  \\
2 & & 2011 JUN 20  & 2011 JUN 28  & 2011 SEP 15  & 2010 DEC 23  \\
3 & & 2011 AUG 20  & 2011 SEP 05  & 2011 SEP 26  & 2011 FEB 16  \\
4 & & 2011 OCT 02  & 2011 OCT 20  & 2012 MAR 24  & 2011 MAR 28  \\
5 & & 2011 NOV 28  & 2011 DEC 06  &              & 2011 MAY 25  \\
6 & & 2012 MAR 22  & 2012 APR 04  &              & 2011 SEP 25  \\
7 & &              &              &              & 2011 DEC 31  \\
8 & &              &              &              & 2012 FEB 16  \\
9 & &              &              &              & 2012 MAR 26  \\
10 & &             &              &              & 2012 MAY 13  \\
11 & &             &              &              & 2012 SEP 23  \\

\enddata
\tablecomments{Scheduling information for the VLBA observations.}

\end{deluxetable}

\begin{deluxetable}{llllrcccc}
\tabletypesize{\scriptsize}
\tablecaption{Source Information\label{tabsou}}
\tablewidth{0pt}
\tablehead{
\colhead{Source} & \colhead{R.A.~(J2000)} & \colhead{Decl.~(J2000)} & \colhead{$\theta_{\rm sep}$} & \colhead{P.A.} &
\colhead{HPBW} & \colhead{F$_{\rm peak}$} & Image rms &\colhead{V$_{\rm LSR}$} \\
\colhead{ }       & \colhead{(h m s)}       & \colhead{($\degr$ ' '')}    & \colhead{($\degr$)}     & \colhead{($\degr$)} &
\colhead{$ \rm mas \times mas \ at \ \degr$} & \colhead{(Jy beam$^{-1}$)} & (Jy beam$^{-1}$) &\colhead{(\kms)} \\}

\startdata

\textbf{G010.62$-$00.38}        & 18 10 28.5629 & $-$19 55 48.738 &     &     & $1.2 \times 0.6 \ at -1.8\degr$ & 42.94 & 0.05 &  $-2.8$  \\
J1808$-$1822                   & 18 08 55.5154 & $-$18 22 53.396 & 1.6 & $-$13 & $1.2 \times 0.6 \ at -2.0\degr$ & 0.011 & 0.001 &  \\
J1821$-$2110                   & 18 21 05.4692 & $-$21 10 45.262 & 2.8 & +117 & $1.3 \times 0.6 \ at -1.3\degr$ & 0.016 & 0.001 &  \\
J1751$-$1950\tablenotemark{a}  & 17 51 41.3438 & $-$19 50 47.506 & 4.4 & $-$90 & $1.2 \times 0.8 \ at +4.4\degr$ & 0.049 & 0.002 &  \\
J1809$-$1520                   & 18 09 10.2094 & $-$15 20 09.699 & 4.6 & $-$4  & $1.2 \times 0.6 \ at -2.3\degr$ & 0.013 & 0.001 &  \\

\textbf{G010.47$+$00.02}        & 18 08 38.2302 & $-$19 51 50.253 &     &     & $ 1.3 \times 0.4  \ at -17.1\degr$ & 22.55 & 0.04 & $+89.9$ \\
J1808$-$1822                   & 18 08 55.5154 & $-$18 22 53.396 & 1.5 & +4 & $ 1.0  \times  0.4  \ at  -12.0 \degr$ & 0.006 & 0.001 &  \\
J1751$-$1950\tablenotemark{a}  & 17 51 41.3438 & $-$19 50 47.506 & 3.2 & $-$90 & $ 1.3  \times  0.4  \ at  -9.8 \degr$ & 0.009 & 0.001 &  \\
J1821$-$2110                   & 18 21 05.4692 & $-$21 10 45.262 & 4.0 & +66 & $ 1.2  \times  0.4  \ at  -9.4 \degr$ & 0.006 & 0.001 &  \\

\textbf{G012.02$-$00.03}        & 18 12 01.8455 & $-$18 31 55.759 &     &      & $ 2.3 \times 1.1 \ at  -0.7\degr$  & 2.37 & 0.01 &  $+108.8$  \\
J1808$-$1822                   & 18 08 55.5154 & $-$18 22 53.396 & 0.8 & $-$78  & $ 4.8 \times 1.9 \ at -9.2\degr$ & 0.039 & 0.001 &  \\
J1809$-$1520                   & 18 09 10.2094 & $-$15 20 09.699 & 3.3 & $-$12  & $ 3.3 \times 1.8 \ at +6.9\degr$ & 0.039 & 0.001 &  \\
J1825$-$1718\tablenotemark{a}  & 18 25 36.5323 & $-$17 18 49.848 & 3.4 & +70  & $ 4.8 \times 1.9 \ at -8.4\degr$ & 0.174 & 0.001 &  \\
J1751$-$1950\tablenotemark{a}  & 17 51 41.3438 & $-$19 50 47.506 & 5.0 & $-$106 & $ 5.2 \times 2.0 \ at -7.6\degr$ & 0.068 & 0.001 &  \\

\textbf{G023.70$-$00.19}        & 18 35 12.3645 & $-$08 17 39.396 &     &     & $ 2.9  \times  1.3  \ at  -5.2\degr$ & 3.99 & 0.02 & $+77.5$ \\
J1825$-$0737\tablenotemark{a}  & 18 25 37.6096 & $-$07 37 30.013 & 2.5 & $-$74 & $ 3.3  \times  1.8  \ at  +3.8\degr$ & 0.315 & 0.003 &  \\
J1846$-$0651\tablenotemark{a}  & 18 46 06.3002 & $-$06 51 27.748 & 3.1 & +63 & $ 3.3  \times  1.8  \ at  +2.2\degr$ & 0.043 & 0.001 &  \\
J1821$-$0502\tablenotemark{a}  & 18 21 11.8094 & $-$05 02 20.087 & 4.8 & $-$47 & $ 3.2  \times  1.8  \ at  +11.7\degr$ & 0.190 & 0.002 &  \\

\enddata
\tablecomments{Positions and source properties for the target maser and the QSO calibrators from the first epoch data.
The peak position of the phase-reference maser channels No.~138, 70, 128, and~131 (for sources from the top to the bottom, respectively)
were calibrated with the ICRF sources and are accurate to within $\pm 2$~mas. For calibrators, we report positions used at
the VLBA correlator. Angular offsets ($\theta_{sep}$) and position angles (P.A.) east of north relative to the maser source are
indicated in columns 4 and 5. Columns 6, 7, and~8 give the natural restoring beam sizes (HPBW), the peak intensities
(F$_{\rm peak}$), and image rms noise of the phase-reference maser channels (at V$_{\rm LSR}$), K-band, and U-band background
sources.} \tablenotetext{a}{Sources belonging to the ICRF catalog.}
\end{deluxetable}

\begin{deluxetable}{llllcll}
\tabletypesize{\scriptsize}
\tablecaption{Parallax \& Proper Motion Fits\label{tabres}}
\tablewidth{0pt}
\tablehead{
\colhead{Maser V$_{\rm LSR}$}   & \colhead{$\Delta x$ \& $\Delta y$ }  & \colhead{Background} & \colhead{Parallax} & D &
\colhead{$\mu_{\rm x}$} & \colhead{$\mu_{\rm y}$} \\
\colhead{(\kms)}  &  \colhead{(mas)}  & \colhead{Source}     & \colhead{(mas)}  & \colhead{(kpc)}  & \colhead{(mas yr$^{-1}$)} &
\colhead{(mas yr$^{-1}$)} \\}

\startdata
\multicolumn{6}{c}{\textbf{G010.47$+$00.02}} \\

$+75.5$ & $+3.6; -329.1$ &  J1751$-$1950 & $0.116 \pm 0.007$ & & $-4.029 \pm 0.014$ & $-6.515 \pm 0.119$  \\ 
$+68.0$ & $-46.0; +68.8$ &  J1751$-$1950 & $0.115 \pm 0.007$ & & $-3.933 \pm 0.012$ & $-6.187 \pm 0.146$  \\ 
$+67.1$ & $-66.7; +16.4$ &  J1751$-$1950 & $0.122 \pm 0.015$ & & $-3.890 \pm 0.039$ & $-6.384 \pm 0.036$  \\ 
$+67.1$ & $-65.3; +69.4$ &  J1751$-$1950 & $0.110 \pm 0.007$ & & $-3.959 \pm 0.019$ & $-6.855 \pm 0.040$  \\ 
$+62.9$ & $-66.7; +89.5$ &  J1751$-$1950 & $0.125 \pm 0.010$ & & $-4.044 \pm 0.028$ & $-6.729 \pm 0.049$  \\ 
$+59.5$ & $+33.3; +57.7$ &  J1751$-$1950 & $0.111 \pm 0.007$ & & $-3.533 \pm 0.018$ & $-6.402 \pm 0.056$  \\ 
$+59.5$ & $-92.1;-136.3$ &  J1751$-$1950 & $0.117 \pm 0.011$ & & $-3.787 \pm 0.030$ & $-6.479 \pm 0.082$  \\ 
    &     &     &                   &                     &                    \\
$+75.5$ & &  Combined & $0.117 \pm 0.008$ & $8.55^{+0.63}_{-0.55}$   &   $-4.028 \pm 0.016$ & $-6.514 \pm 0.089$  \\
$+68.0$ & &           &                   &                          &   $-3.932 \pm 0.013$ & $-6.185 \pm 0.073$  \\
$+67.1$ & &           &                   &                          &   $-3.890 \pm 0.022$ & $-6.385 \pm 0.122$  \\
$+67.1$ & &           &                   &                          &   $-3.959 \pm 0.022$ & $-6.855 \pm 0.122$  \\
$+62.9$ & &           &                   &                          &   $-4.044 \pm 0.022$ & $-6.729 \pm 0.122$  \\
$+59.5$ & &           &                   &                          &   $-3.534 \pm 0.022$ & $-6.402 \pm 0.122$  \\
$+59.5$ & &           &                   &                          &   $-3.788 \pm 0.022$ & $-6.480 \pm 0.122$  \\
\multicolumn{6}{c}{\textbf{Best-value for the Secular proper motion of G010.47$+$00.02}} \\
    & &          &                   &  &                       &                    \\
    & &          & $0.117 \pm 0.008$ & $8.55^{+0.63}_{-0.55}$   & $-3.860 \pm 0.015 $ &  $-6.403 \pm 0.076 $  \\
    & &          &                   &  &                       &                    \\
\multicolumn{6}{c}{\textbf{G010.62$-$00.38}} \\

$-14.2$ & $+607.0;-605.1$ &  J1751$-$1950 & $0.191 \pm 0.034$ & & $-0.632 \pm 0.084$ & $-0.503 \pm 0.122$  \\
$-14.2$ & &  J1821$-$2110 & $0.200 \pm 0.040$ &                    &                     \\
$-1.1$  & $+386.9;-247.3$ &  J1751$-$1950 & $0.202 \pm 0.034$ & & $-0.427 \pm 0.084$ & $-1.474 \pm 0.084$  \\
$-1.1$  & &  J1821$-$2110 & $0.216 \pm 0.026$ &                    &                     \\
$+1.0$  & $+324.8;-194.5$ &  J1751$-$1950 & $0.200 \pm 0.030$ & & $-0.355 \pm 0.075$ & $-0.876 \pm 0.053$  \\
$+1.0$  & &  J1821$-$2110 & $0.213 \pm 0.032$ &                    &                     \\
    &         &                   &                     &                    \\
$-14.2$ & &  Combined & $0.202 \pm 0.019$ & $4.95^{+0.51}_{-0.43}$ & $-0.632 \pm 0.084$ & $-0.503 \pm 0.122$  \\
$-1.1$  & &          &                    &  & $-0.427 \pm 0.084$ & $-1.474 \pm 0.084$  \\
$+1.0$  & &          &                    &  & $-0.355 \pm 0.075$ & $-0.876 \pm 0.053$  \\
\multicolumn{6}{c}{\textbf{Best-value for the Secular proper motion of G010.62$-$00.38}} \\
    & &          &                   &  &                     &                    \\
    & &          & $0.202 \pm 0.019$ & $4.95^{+0.51}_{-0.43}$ & $-0.366 \pm 0.081$  &  $-0.600 \pm 0.055 $  \\
    & &          &                   &  &                     &                    \\
\multicolumn{6}{c}{\textbf{G012.02$-$00.03}} \\

$+108.0$ & +0.6; +1.0 &  J1808$-$1822 & $0.098 \pm 0.001$ & & $-4.205 \pm 0.003$ & $-7.963 \pm 0.349$  \\
$+108.8$ & 0.0; 0.0   &  J1808$-$1822 & $0.115 \pm 0.008$ & & $-4.008 \pm 0.023$ & $-7.551 \pm 0.279$  \\

$+108.0$ & &  Combined & $0.106 \pm 0.008$ & $9.43^{+0.77}_{-0.66}$ & $-4.205 \pm 0.022$ & $-7.963 \pm 0.265$  \\
$+108.8$ & &           &                   &                        & $-4.008 \pm 0.022$ & $-7.551 \pm 0.265$  \\
    & &          &                   &                     &                    \\
\multicolumn{6}{c}{\textbf{G023.70$-$00.19}} \\

$+77.5$  & 0.0; 0.0      &  J1825$-$0737 & $0.163 \pm 0.038$ & & $-3.226 \pm 0.096$ & $-6.704 \pm 0.224$  \\ 
$+77.5$  &               &  J1846$-$0651 & $0.187 \pm 0.028$ & & $-3.100 \pm 0.071$ & $-6.237 \pm 0.115$  \\
$+76.7$  & $+2.7; -21.5$ &  J1825$-$0737 & $0.154 \pm 0.034$ & & $-3.299 \pm 0.084$ & $-6.518 \pm 0.256$  \\ 
$+76.7$  &               &  J1846$-$0651 & $0.173 \pm 0.041$ & & $-3.171 \pm 0.103$ & $-6.046 \pm 0.167$  \\
$+79.0$  & $+1.6; -2.5$  &  J1825$-$0737 & $0.136 \pm 0.041$ & & $-3.288 \pm 0.103$ & $-6.577 \pm 0.231$  \\ 
$+79.0$  &               &  J1846$-$0651 & $0.161 \pm 0.036$ & & $-3.159 \pm 0.090$ & $-6.104 \pm 0.116$  \\
    & &          &                   &        &             &                    \\
$+77.5$  & &  Combined & $0.161 \pm 0.024$ & $6.21^{+1.0}_{-0.80}$ & $-3.164 \pm 0.059$ & $-6.474 \pm 0.136$  \\
$+76.7$  & &          &                    &                       & $-3.235 \pm 0.059$ & $-6.283 \pm 0.136$  \\
$+79.0$  & &          &                    &                       & $-3.223 \pm 0.059$ & $-6.338 \pm 0.136$  \\

\enddata
\tablecomments{Columns~1 and~2 report the LSR velocity and the offset positions of the reference maser spots, respectively,
with respect to the phase-reference maser channels in Table~\ref{tabsou}. Column 3 indicates the background sources whose data were
used to model the relative proper motion of the maser for the parallax fit; column 4 and 5 report the fitted parallax and distance;
columns 6 and 7 give the fitted proper motions along the east and north directions, respectively. Combined fit used a single
parallax parameter for the maser spots relative to the QSOs. Quoted uncertainties for individual parallax fits are from the formal
fitting uncertainty.
The Secular proper motion of G010.47$+$00.02 and G010.62$-$00.38 was obtained by the combined fit value corrected for the internal
proper motions of the overall H$_2$O maser distribution.}

\end{deluxetable}

\begin{deluxetable}{lrrrrcrrrcl}
\tabletypesize{\scriptsize}
\rotate
\tablecaption{Galactic motion of sources in the inner Milky Way\label{tabinner}}
\tablewidth{0pt}
\tablehead{
\colhead{Source} & \colhead{$\ell$}  & \colhead{b} & \colhead{V$_{\rm LSR}$}& \colhead{D} & \colhead{R} & \colhead{U$_s^{\rm G.C.}$} & \colhead{V$_s^{\rm G.C.}$}  & \colhead{W$_s^{\rm G.C.}$} & Ref. & Arm \\
 & \colhead{(deg)} & \colhead{(deg)}& \colhead{(km s$^{-1}$)}& \colhead{(kpc)} & \colhead{(kpc)} & \colhead{(km s$^{-1}$)}& \colhead{(km s$^{-1}$)}& \colhead{(km s$^{-1}$)} &  & \\}

\startdata

G009.62$+$00.19 & 9.621  & $+0.196$ & $+5.0 \pm 3.1$ & $5.15^{+0.77}_{-0.66}$ & $3.4^{+0.6}_{-0.7}$ & $-36.1 \pm 16.7$  & $+191.9 \pm 15.1 $ &
$-10.1 \pm 4.1$  & 1 &  Norma \\
G010.47$+$00.02 & 10.472 & $+0.027$ & $+68.9 \pm 4.5$  & $8.55^{+0.63}_{-0.55}$ & $1.6^{+0.2}_{-0.1}$ & $+30.3 \pm 21.5$  & $+121.7 \pm 15.8 $ &
$+18.2 \pm 1.8$  & 2 & Connecting\\
G010.62$-$00.38 (W31) & 10.624 & $-0.383$ & $-3.0 \pm 2.7$  & $4.95^{+0.51}_{-0.43}$ & $3.6^{+0.4}_{-0.4}$ & $-60.8 \pm 14.4$ & $+228.2 \pm  6.7 $ &
$ +8.1 \pm 1.8$  & 2 & near 3~kpc\\
G012.02$-$00.03 & 12.025 & $-0.031$ & $+109.8 \pm 2.4$  & $9.43^{+0.77}_{-0.66}$ & $2.1^{+0.5}_{-0.3}$ & $+25.5 \pm 32.7$  & $+215.0 \pm 26.8 $ &
$ +1.5 \pm 5.8$  & 2 & far 3~kpc\\
G023.44$-$00.18 & 23.440 & $-0.182$ & $+104.2 \pm 4.3$  & $5.88^{+1.37}_{-0.93}$ & $3.8^{+0.5}_{-0.4}$ & $ +4.7 \pm 42.7$  & $+225.4 \pm 18.1 $ &
$ +2.0 \pm 3.1$  & 3 & Norma \\
G023.70$-$00.19 & 23.707 & $-0.198$ & $+76.5 \pm 10$  & $6.21^{+1.0}_{-0.80}$  & $3.7^{+0.4}_{-0.3}$ & $+51.4 \pm 15.5$  & $+167.1 \pm 12.0 $ &
$ +4.6 \pm 2.6$  & 2 & Norma \\

\enddata
\tablecomments{Proper motion of sources that belong to the inner Milky Way with respect to a reference frame joined with the Galactic
center (see Section~\ref{distance_results}). For converting the heliocentric measurements to the Galactic center
reference frame, we considered the best estimates of R$_0 =8.38$~kpc and $\Theta_0=243$~\kms\ given in \citet{Reid2013} and the updated
values of the Solar motion in \citet{Schoenrich2010}. Column 4 gives the LSR velocity of the regions assumed in the calculation, as
inferred from the CS\,(2--1) line survey by \citet{Bronfman1996} with the exception of G010.62$-$00.38 and G023.70$-$00.19
(see Appendix~\ref{appendix}).
Columns 5 and 6 report the heliocentric and projected Galactocentric distance of each source, respectively. Columns 7, 8, and 9 give
the velocity components toward the Galactic center, in the direction of Galactic rotation, and toward the North Galactic
Pole, respectively. Column~10 lists the reference papers for the trigonometric parallax measurements: (1) \citet{Sanna2009}; (2)
this work; (3) \citet{Brunthaler2009}. Last column lists the Galactic arms associated with the HMSFRs as presented in Section~\ref{association}.}

\end{deluxetable}

\end{document}